\documentclass[10pt]{iopart}
\input{epsf}
\usepackage{iopams}

\begin{document}
\title{Heisenberg picture operators in the quantum state diffusion model}
\author{ Heinz--Peter Breuer, Bernd Kappler, Francesco Petruccione}

\address{Albert-Ludwigs-Universit\"at, Fakult\"at f\"ur Physik, \\
Hermann-Herder Stra{\ss}e 3, D--79104 Freiburg im Breisgau,
Federal Republic of Germany}
\begin{abstract}
A stochastic simulation algorithm for the computation of multitime
correlation functions which is based on the quantum state diffusion
model of open systems is developed. The crucial point of the proposed
scheme is a suitable extension of the quantum master equation to a
doubled Hilbert space which is then unraveled by a stochastic
differential equation.
\end{abstract}
\pacs{42.50.Lc,02.70.Lq}
\maketitle

%---------------------------------------
%  INRODUCTION
%---------------------------------------
Within the framework of the recently developed stochastic wave
function approach to open quantum systems
\cite{MolmerPRL68,Carmichael,GardinerPRA46,ZollerPRA46,Gisin:92,Gisin:93,BP:QS3,BP:QS4}
the state of a system is not described by a reduced density matrix but
by a pure stochastic state vector $\psi_t$ whose covariance matrix is
equal to the reduced density matrix of the system.  One of these
models which was motivated by a dynamical description of the
measurement process \cite{Gisin:84} is the quantum state diffusion
model introduced by Gisin and Percival \cite{Gisin:92,Gisin:93}. In
this approach, the time-evolution of the wave function $\psi_t$ is
governed by the Ito stochastic differential equation
\begin{eqnarray}
  \label{sde_eq}
  d\psi_t=&-&\mathrm{i}{H}\psi_t d t+\frac{1}{2}\sum_j
  \left[2\langle {L}_j^\dagger\rangle_{\psi_t}
  {L}_j-{L}_j^\dagger {L}_j
  -\langle {L}_j\rangle_{\psi_t}
  \langle {L}_j^\dagger\rangle_{\psi_t}\right]\psi_t d t 
  \nonumber\\
  &+&\sum_j\left[{L}_j-
  \langle {L}_j\rangle_{\psi_t}\right]\psi_td\xi_{jt},
\end{eqnarray}
where $\langle {L}_j\rangle_{\psi_t}$ is a short-hand notation 
for $\langle\psi_t| {L}|\psi_t\rangle$, and $d\xi_{jt}$ is
the differential of a complex valued Wiener process with means and 
correlations 
\begin{equation}
  \label{wien_eq}
  \langle d\xi_{jt}\rangle=\langle  d\xi_{it}d\xi_{jt}\rangle
  =0,\quad \langle d\xi_{it} d\xi_{jt}^*\rangle=\delta_{ij} d t.
\end{equation}
The operators $H$ and $L_j$ acting in the Hilbert space $\mathcal H$
of the system are the free Hamiltonian and the Lindblad operators
describing dissipation, respectively.  The link to the density matrix
description of open quantum systems is established -- as mentioned
above -- through the covariance matrix of the stochastic wave function
$\psi_t$, i.~e.,
\begin{equation}
  \label{rho_psi_eq}
  \rho_t=\mathrm{E} \left(|\psi_t\rangle\langle\psi_t|\right).
\end{equation}
The symbol E denotes the expectation value with respect of the stochastic
processes $\psi_t$.  The equation of motion of the density matrix
$\rho_t$ is obtained by inserting eq.~(\ref{sde_eq}) in
eq.~(\ref{rho_psi_eq}) which yields the quantum master equation
\begin{equation}
  \label{QME_eq}
  \dot\rho(t)=-{\mathrm i}\left[H, \rho(t)\right]
  +\frac{1}{2}\sum_j\left[
  2L_j\rho(t) L_j^\dagger-L_j^\dagger L_j\rho(t)-\rho(t)L_j^\dagger L_j\right].
\end{equation}
This equation -- or alternatively the stochastic differential equation
(\ref{sde_eq}) -- determines the time-evolution of one-time
expectation values of system observables. Multitime correlation
functions which are of special interest in quantum optics or in solid
state physics are not specified by these equations. In order to define
these quantities, we will first define the matrix elements of some
system operator $A$ in the Heisenberg picture.  In the density matrix
approach, these matrix elements are defined through the quantum
regression theorem \cite{GardinerQN,Walls} as
\begin{equation}
  \label{heis_def_eq}
  A_t(\phi_0,\psi_0) \equiv \langle\phi_0,t_0|A(t)|\psi_0,t_0
  \rangle =\mbox{Tr}\Big\{AV(t,t_0)\left\{|\psi_0\rangle
  \langle\phi_0|\right\}\Big\},
\end{equation}
where $V(t,t_0)$ is the time-evolution superoperator corresponding to
the quantum master equation (\ref{QME_eq}).  Unfortunately, the
quantum regression theorem cannot be applied directly to the
stochastic wave function approach, since the initial ``density
matrix'' $|\psi_0\rangle\langle\phi_0|$ is not necessarily Hermitian,
and hence it can in general not be the covariance matrix of some
stochastic wave function. This problem can be resolved by
extending the quantum master equation into a doubled Hilbert space
$\tilde{\mathcal H} ={\mathcal H}\oplus {\mathcal H}$ in the following
way: we define a density matrix $\widetilde{\rho}(t)$ as
\begin{equation}
  \label{rho_tilde_eq}
  \widetilde{\rho}(t)=\left(
  \begin{array}{cc}
  \widetilde{\rho}_{11}(t)&\widetilde{\rho}_{12}(t)\\
  \widetilde{\rho}_{21}(t)&\widetilde{\rho}_{22}(t)\end{array}\right),
\end{equation}
where $\widetilde{\rho}_{ij}(t)$ are operators on ${\mathcal H}$ and
accordingly replace the Hamiltonian $H$ and the Lindblad operators
$L_j$ by the operators 
\begin{equation}
\label{H_L_ext_eq}
  \widetilde{H}=
  \left(\begin{array}{cc}
  H&0\\
  0&H\end{array}\right),\quad
  \widetilde{L}_j=
  \left(\begin{array}{cc}
  L_j&0\\
  0&L_j\end{array}\right)
\end{equation}
in the doubled Hilbert space $\widetilde{\mathcal H}$. Then we formulate the
extended quantum master equation
\begin{equation}
  \label{ext_QME_eq}
  \dot{\widetilde\rho}(t)=-\mathrm{i}\left[\widetilde{H}, 
  \widetilde\rho(t)\right]
  +\frac{1}{2}\sum_j\left[
  2\widetilde{L}_j\widetilde\rho(t) \widetilde{L}_j^\dagger-
  \widetilde{L}_j^\dagger \widetilde{L}_j\widetilde\rho(t)-
  \widetilde\rho(t) \widetilde{L}_j^\dagger \widetilde{L}_j\right].
\end{equation}
The crucial point of this construction is that each element
$\widetilde{\rho}_{ij}(t)$ of the density matrix $\widetilde{\rho}(t)$
is a solution of the original quantum master equation
(\ref{QME_eq}). Consider now the initial condition
\begin{equation}
  \label{ext_init_eq}
  \widetilde\rho(t_0)=|\theta_0\rangle\langle\theta_0|\equiv\frac{1}{2}\left(
  \begin{array}{cc}
  |\phi_0\rangle\langle\phi_0|&|\phi_0\rangle\langle\psi_0|\\
  |\psi_0\rangle\langle\phi_0|&|\psi_0\rangle\langle\psi_0|\end{array}\right),
\end{equation}
where $\theta_0\equiv(\phi_0,\psi_0)^{\mathrm T}/\sqrt{2}$ is an
element of the doubled Hilbert space $\widetilde{\mathcal H}$.
(Throughout this letter the superscript ${\mathrm T}$ denotes the
transpose).  Obviously, the matrix elements of some operator $A$ are
then given by
\begin{equation}
  \label{Heis_ext_eq}
  A_t(\phi_0,\psi_0)=
  2\mbox{Tr}\Big\{A\widetilde{\rho}_{21}(t)\Big\}.
\end{equation}
By construction, the initial density matrix $\widetilde\rho(t_0)=
|\theta_0\rangle\langle\theta_0|$ is positive and we may choose any
unraveling of the extended quantum master equation (\ref{ext_QME_eq})
by a stochastic process for the calculation of its
time-evolution and hence for the calculation of operators in the
Heisenberg picture (A similar idea has been proposed in
Ref. \cite{ZollerPRA46} Appendix D). 

\begin{figure}[p]
  \begin{center}
    \leavevmode  \epsfxsize .8\linewidth\epsffile{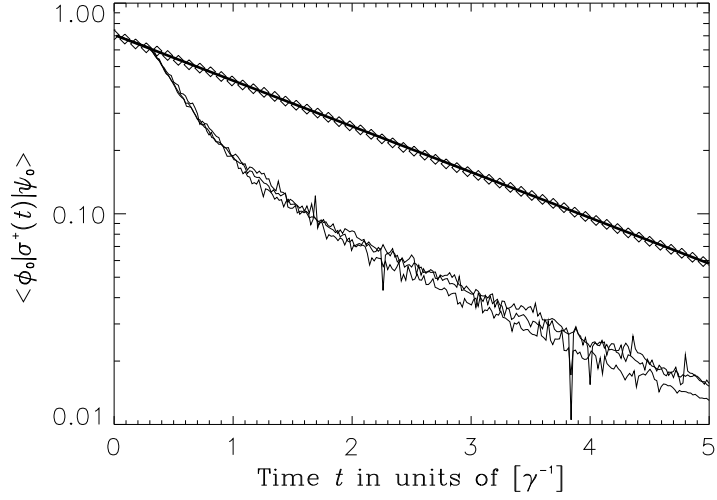}  
    \vspace{2ex}
    \caption{Calculation of Heisenberg operator matrix element 
        $\langle\phi_0|\sigma^+(t)|\phi_0\rangle$: analytical solution 
        (thick line),  numerical solution using the quantum state
        diffusion unraveling of the extended quantum master equation for
        $10^3$ realizations  (diamonds), and the method proposed by Gisin 
        (thin lines) for the step-sizes $h=0.01,0.001,0.0001$.}
\label{fig1}
\end{center}
\end{figure}

Applying the above procedure to the quantum state
diffusion model we obtain for example the equation of motion for the
wave function $\theta_t=(\phi_t,\psi_t)^{\mathrm
T}\in\widetilde{\mathcal H}$ in the Ito form
\begin{eqnarray}
  \label{ext_sde_eq}
  d\theta_t=&-&\mathrm{i}\widetilde{H}\theta_t dt+\frac{1}{2}\sum_j
  \left[2\langle \widetilde{L}_j^\dagger\rangle_{\theta_t}
  \widetilde{L}_j-\widetilde{L}_j^\dagger \widetilde{L}_j
  -\langle \widetilde{L}_j\rangle_{\theta_t}
  \langle \widetilde{L}_j^\dagger\rangle_{\theta_t}\right]\theta_t dt 
  \nonumber\\
  &+&\sum_j\left[\widetilde{L}-
  \langle \widetilde{L}_j\rangle_{\theta_t}\right]\theta_td\xi_{jt}.
\end{eqnarray}
The matrix elements of $A$ are simply obtained as
\begin{equation}
  \label{mat_el_eq}
  A_t(\phi_0,\psi_0)=2\mathrm{E}_{\theta_0}(\langle\phi_t|A|\psi_t\rangle),
\end{equation}
where $\mathrm{E}_{\theta_0}$ denotes the expectation value with
respect to the initial condition $\theta_0$.  Note, that
eq.~(\ref{ext_sde_eq}) is constructed in such a way that the norm of
the state vector $\theta_t$ is preserved, i.~e.,
$||\theta_t||^2=||\phi_t||^2+||\psi_t||^2=1$. From a numerical point
of view it is more efficient to drop this restriction and to work with
unnormalized state vectors $\hat\theta_t$, whose time-evolution is
governed by the quasi-linear stochastic differential equation
\cite{Gisin:92}
\begin{equation}
  \label{ql_ext_sde_eq}
  d\hat\theta_t=-\mathrm{i}\widetilde{H}\hat\theta_t dt
  +\sum_j\widetilde{L}_j\hat\theta_t\left(d\xi_{jt}
  +\langle\widetilde{L}_j^\dagger\rangle_{\theta_t}dt\right)
  -\frac{1}{2}\sum_j\widetilde{L}_j^\dagger \widetilde{L}_j\hat\theta_tdt.
\end{equation}
Accordingly, the matrix elements of the operator $A$ are defined as 
\begin{equation}
  \label{ql_mat_el_eq}
  A_t(\phi_0,\psi_0)=2E_{\theta_0}\left(\langle\hat\phi_t|A|\hat\psi_t\rangle
    /||\hat\theta_t||^2\right).
\end{equation}
As a particular example, we consider a two-level system with $H=0$
coupled to the vacuum using the Lindblad operator $\sigma^-$, and
calculate the matrix element
$\langle\phi_0|\sigma^+(t)|\psi_0\rangle$, where
$\phi_0=(1,0)^{\mathrm T}$ and $\psi_0=(1,1)^{\mathrm
T}/\sqrt{2}$. The analytical solution
\begin{equation}
  \label{ana_sol_eq}
  \langle\phi_0|\sigma^+(t)|\psi_0\rangle=\frac{1}{\sqrt{2}}e^{-t/2}
\end{equation}
is readily obtained by integrating the quantum master equation. In
Fig.\ref{fig1} we compare the numerical solution obtained using the
scheme described above for $10^3$ realizations (diamonds) with the
analytical solution (thick line). Obviously, both solutions are in
excellent agreement.
 
Alternatively, Gisin proposes in Ref.~\cite{Gisin:Heis} a similar
scheme for the calculation of matrix elements which is based on the
coupled system of stochastic differential equations
\begin{eqnarray}
  \label{phi_psi_eq}
  \lefteqn{\hspace*{-4em}d\psi_t=-\mathrm{i}H\psi_t dt+\frac{1}{2}\sum_j
  \left[2l_j(\psi_t,\phi_t)^*L_j-L_j^\dagger L_j
  -l_j(\phi_t,\psi_t)l_j(\psi_t,\phi_t)^*\right]\psi_t dt }
  \nonumber\\ 
  &+&\sum_j\left[L-l_j(\phi_t,\psi_t)\right]\psi_td\xi_{jt},\nonumber\\
  \lefteqn{\hspace*{-4em}d\phi_t=-\mathrm{i}H\phi_t dt+\frac{1}{2}\sum_j
  \left[2l_j(\phi_t,\psi_t)^*L_j-L_j^\dagger L_j
  -l_j(\psi_t,\phi_t)l_j(\phi_t,\psi_t)^*\right]\phi_t dt }
  \nonumber\\ 
  &+&\sum_j\left[L-l_j(\psi_t,\phi_t)\right]\phi_td\xi_{jt},
\end{eqnarray}
where $l_j(\alpha,\beta)=\langle\alpha|L_j|\beta\rangle
/\langle\alpha|\beta\rangle$. These equations are constructed in such
a way that the scalar product $\langle\phi_t|\psi_t\rangle$ remains
constant during the time-evolution of the system, i.~e., the matrix
element of the unity operator $I_t=I$ are calculated correctly for
each realization of the stochastic process (and not only in the
mean). In addition, he also proposes in Ref.~\cite{Gisin:Heis} a pair
of quasi-linear equations, which could be used for the numerical
simulation. However, although the above equations correctly reproduce
the equation of motion for the matrix elements, the numerical
integration of the stochastic differential equations for the system
described above, suggests that these equations are not stable in
general. In order to demonstrate this, we have also plotted in
Fig.~\ref{fig1} the numerical solution of the quasi-linear stochastic
differential equations for various step-sizes ($h=0.01,0.001,0.0001$)
and $10^4$ realization each. The systematic deviation of the numerical
and analytical solutions for $t\gtrsim 0.3\gamma^{-1}$ is evident. We
believe, that these deviations are due to the fact, that the solution
of the deterministic part of the stochastic differential equation is
unstable for this particular model which leads to immense fluctuations
in the solution of the stochastic differential equation. Note, that
the fluctuations are even much larger for the integration of the
``unity-preserving'' equation~(\ref{phi_psi_eq}).

\begin{figure}[b]
  \begin{center}
    \leavevmode  
    \parbox[b]{.49\linewidth}
        {\epsfxsize\linewidth\epsffile{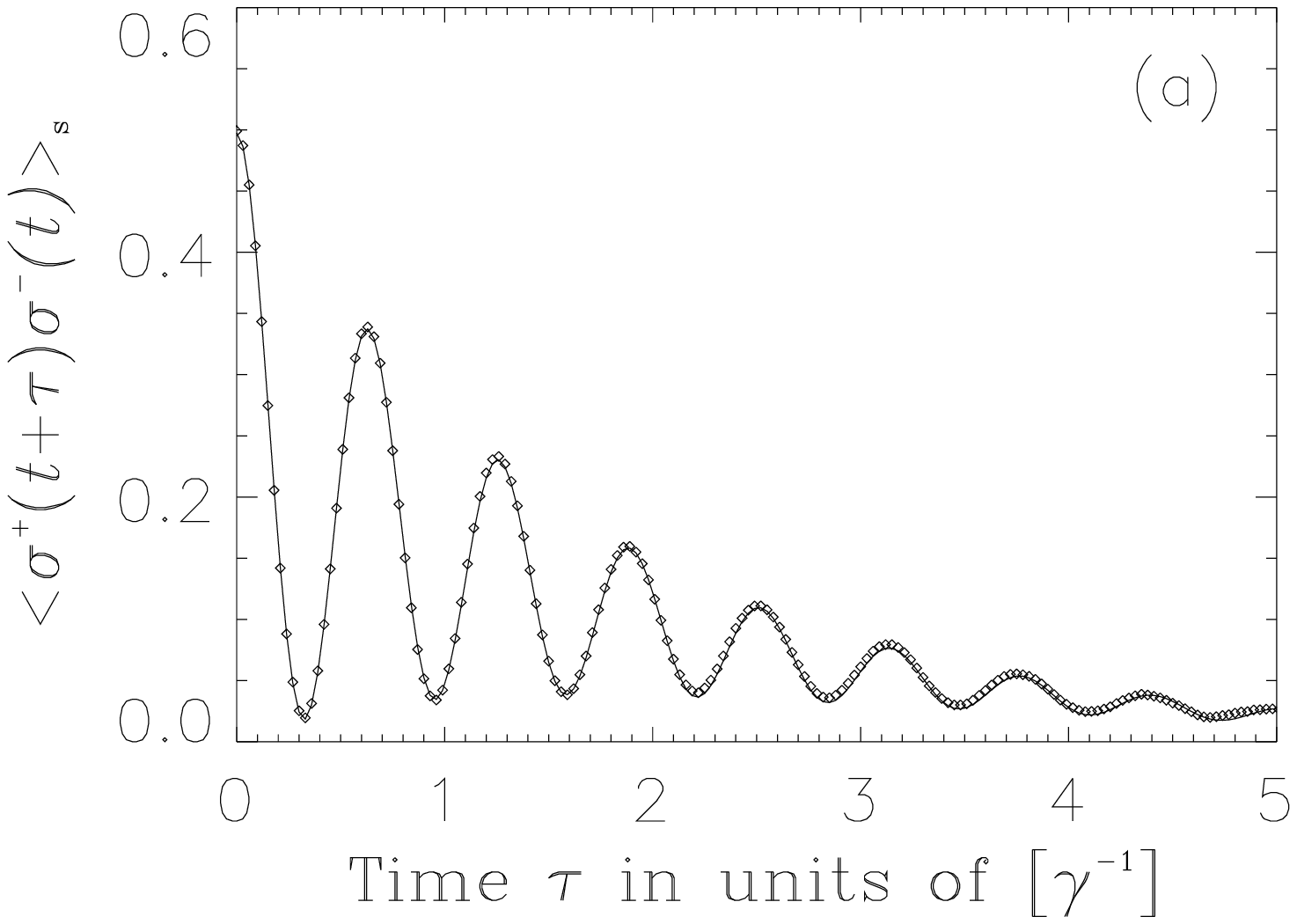}}
        \parbox[b]{.49\linewidth}
        {\epsfxsize\linewidth\epsffile{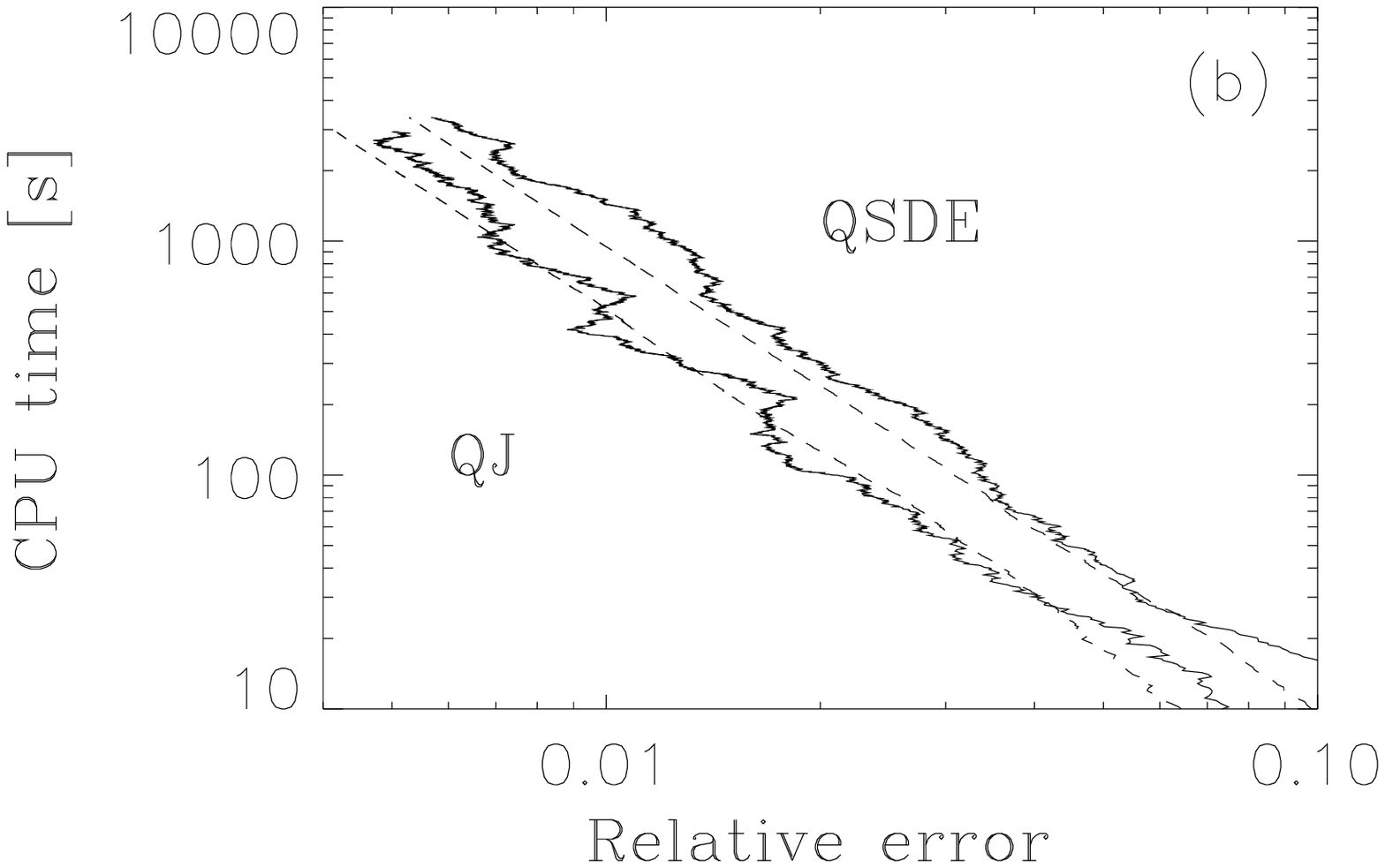}} 
    \vspace{2ex}
    \caption{Calculation of the first order correlation function 
        $\langle \sigma^+(\tau)\sigma^-\rangle_{s}$ for a 
        coherently driven two-level atom on resonance. 
        (a) Analytical solution vs.~the numerical solution (diamonds) 
         using the quantum state diffusion model for $10^4$ realizations.
        (b) CPU time in seconds vs.~the relative error for the simulation 
        using the quantum state diffusion model (QSDE) and the quantum jump 
        method (QJ). 
        The solid lines  represent the mean square deviation of the numerical
         solution from the exact solution and the dashed lines show the 
        estimated standard deviation of the numerical solution.}
\label{fig2}
\end{center}
\end{figure}

The simulation algorithm in the doubled Hilbert space for the
calculation of matrix elements in the Heisenberg picture is the basis
for the computation of multitime correlation functions such as
$g(t,t+\tau)=\langle\psi_0|A(t+\tau)B(t)|\psi_0\rangle$ and we propose
the following procedure: start in the state $\psi_0$ and propagate it
up to the time $t$ using the stochastic differential
equation~(\ref{sde_eq}) to obtain $\psi_t$.  Define the state vector
$\theta_t=(\psi_t,B\psi_t)^{\mathrm T}/\sqrt{1+||B\psi_t||^2}$ and
propagate it up to the time $t+\tau$ by integrating the extended
stochastic differential equation~(\ref{ext_sde_eq}). The two-time
correlation function $g(t,t+\tau)$ is then given by
\begin{equation}
   \label{corr_def_eq}
   g(t,t+\tau) = {\mathrm E}\left[\left(1+||B\psi_t||^2 \right)
   \langle\phi_{t+\tau}|A|\psi_{t+\tau}\rangle\right].
\end{equation}
As a specific example we have computed the first order correlation
function $\langle\sigma^+(t+\tau)\sigma^-(t)\rangle_{s}$ for a
coherently driven two-level atom on resonance in the steady state with
Rabi frequency $\Omega=10\gamma$. To this end, we started with a
random initial state vector $\psi_0$ drawn from a uniform distribution
on ${\mathcal H}$ and propagated it up to $t=30 \gamma^{-1}$ in order
to reach the steady state regime. Then we proceeded as described
above. The result of the numerical simulation is shown in
Fig.~\ref{fig2} (a) for $10^4$ realizations. The numerical performance
of the algorithm is demonstrated in Fig.~\ref{fig2} (b) where we have
plotted the computational time which is necessary to obtain a given
accuracy measured by the relative mean square error (solid line) and
the estimated standard deviation of the samples (dashed line).  These
results are compared with an alternative procedure which is based on
an unraveling of the extended quantum master equation by a piecewise
deterministic jump process (see Ref.~\cite{BP:QS12}).  The
algorithm based on quantum jumps is about two times faster than the
one based on the quantum state diffusion model. At a first glance,
this result is surprising, since the individual realizations of the
diffusion process are smooth and ``closer'' to the real solution. But
this is outweighed by the fact that for the integration of the
stochastic differential equation we have to draw two random numbers
per time step and Lindblad operator, whereas in the quantum jump
method we have to generate only two random numbers per jump. Thus, a
single realization of the diffusion process is more accurate, but
takes longer to be computed.

To summarize, we have shown that operators in the Heisenberg picture
and multitime correlation functions can be calculated within the
framework of the quantum state diffusion model by extending the
stochastic differential equation which governs the time-evolution of
the wave function to the doubled Hilbert space. This procedure is in
complete agreement with the quantum regression theorem. However, we
have also shown that the latter fact is not sufficient to ensure that
a particular simulation algorithm is of practical use: Although the
algorithm proposed in Ref.~\cite{Gisin:Heis} is in accordance with the
quantum regression theorem, it seems not to be stable in general. On the other
hand, the scheme we proposed in this letter completely relies on the
numerical stability of the quantum state diffusion model.

\section*{References}

\bibliographystyle{prsty}  % Phys. Rev. / APS Style

%%----------Fig. 1------------------------------

%%----------Fig. 2------------------------------

\end{document}